\documentclass[aps,prd,nofootinbib,twocolumn]{revtex4}
\pdfoutput=1

\usepackage{graphicx}
\usepackage{hyperref}

\begin{document}

\title{Negative running prevents eternal inflation}

\author{William H.\ Kinney} 
\email{whkinney@buffalo.edu}
\affiliation{Dept. of Physics, University at Buffalo,
        the State University of New York, Buffalo, NY 14260-1500}

\author{Katherine Freese}
\email[]{ktfreese@umich.edu}
\affiliation{
 Department of Physics,
 University of Michigan,
 Ann Arbor, MI 48109}

\date{\today}

\begin{abstract}
Current data from the Planck satellite and the BICEP2 telescope favor, at around the $2 \sigma$ level, negative running of the spectral index of curvature perturbations from inflation. We show that for negative running $\alpha < 0$, the curvature perturbation amplitude has a {\it maximum} on scales larger than our current horizon size. A condition for the absence of eternal inflation is that the curvature perturbation amplitude always remain below unity on superhorizon scales. For current bounds on $n_{\rm S}$ from Planck, this corresponds to an upper bound of the running $\alpha < - 4 \times 10^{-5}$, so that even tiny running of the scalar spectral index is sufficient to prevent eternal inflation from occurring, as long as the running remains negative on scales outside the horizon. In single-field inflation models, negative running is associated with a finite duration of inflation: we show that eternal inflation may not occur even in cases where inflation lasts as long as $10^4$ e-folds.
\end{abstract}


\maketitle

\section{Introduction}

Inflation \cite{Guth:1980zm,Linde:1981mu,Albrecht:1982wi} has emerged as the standard paradigm for modeling the behavior of the very early universe. In addition to explaining the flatness and homogeneity of the cosmos, inflation predicts the generation of perturbations from quantum fluctuations in the early universe \cite{Kazanas:1980tx,Starobinsky:1980te,Sato:1981ds,Sato:1980yn,Mukhanov:1981xt,Mukhanov:2003xw,Linde:1983gd,Hawking:1982cz,Hawking:1982my,Starobinsky:1982ee,Guth:1982ec,Bardeen:1983qw}, a prediction which has been tested to high precision in measurements of the Cosmic Microwave Background (CMB). The temperature anisotropy of the CMB has been measured in exquisite detail by the Planck satellite \cite{Ade:2013kta,Ade:2013zuv,Ade:2013uln}, and recent measurement of the CMB polarization by the BICEP2 telescope has provided clear evidence of primordial gravitational waves consistent with the predictions of inflation \cite{Ade:2014xna}. 

The Planck and BICEP data are consistent with the simplest inflationary models. An example is inflation in a quadratic monomial potential, $V(\phi) \propto \phi^2,$ which predicts a tensor/scalar ratio $r \simeq 0.15$, consistent with BICEP2 constraints, and a scalar spectral index $n_{\rm S} \simeq 0.96$, consistent with Planck constraints. Such ``large-field'' inflationary models have field excursion $\Delta\phi > M_{\rm P}$ during inflation, where $M_{\rm P}$ is the reduced Planck Mass. Such potentials have the interesting property that, for field values $\phi \gg M_{\rm P}$, the amplitude of quantum fluctuations in the field becomes larger than the classical field variation, so that the field is as likely to roll {\it up} the potential as it is to roll down the potential. Therefore, in a statistical sense, inflation never ends: there will always be regions of the universe where the field has fluctuated upward, rather than downward, and inflation becomes a quasi-stationary, infinitely self-reproducing state of eternal inflation \cite{Vilenkin:1983xq,Guth:1985ya,Linde:1986fc,Linde:1986fd}. 

In this paper we consider the question of whether the BICEP2 data {\it imply} eternal inflation. We focus in particular on the fact that Planck + BICEP2 weakly favor a scale-dependent spectral index, a so-called {\it running} of the primordial power spectrum. Such running of the power spectrum would rule out all simple monomial potentials, requiring a more complex (and more finely tuned) inflationary potential. We find that for even a small negative running, of order $\alpha \sim 10^{-4}$, eternal inflation is prevented, even in cases where inflation continues for many e-folds. Therefore, it is premature to conclude that the large tensor signal favored by BICEP2 is consistent only with models leading to eternal inflation. The paper is organized as follows: Section \ref{sec:running} considers the current evidence for running of the spectral index. Section \ref{sec:eternal} discusses the relationship between eternal inflation and the amplitude of the curvature perturbation spectrum. Section \ref{sec:limit} discusses suppression of eternal inflation in the case of negative running of the curvature power spectrum. Section \ref{sec:conclusions} presents a summary and conclusions.  

\section{Evidence for running of the spectral index}
\label{sec:running}

\begin{figure}
\includegraphics[keepaspectratio,width=1.00\columnwidth,height=0.40\textheight]{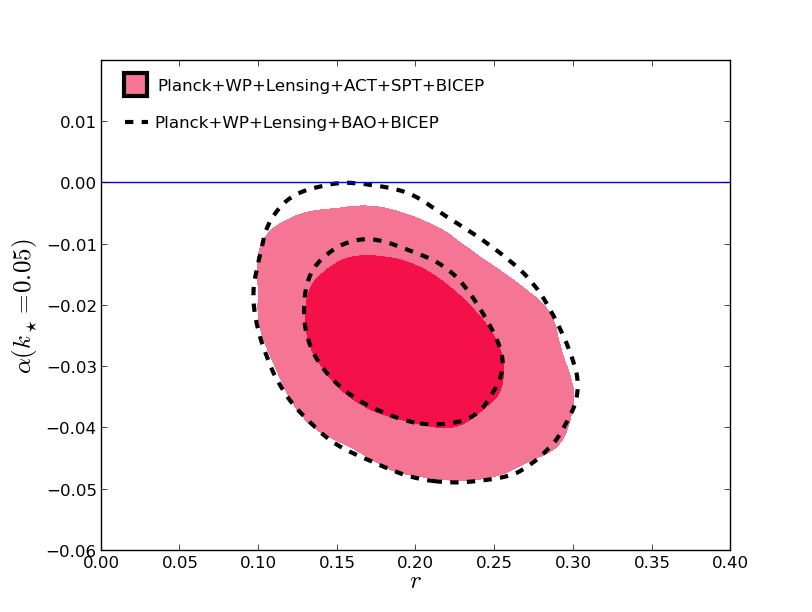}
\caption{Constraints on the tensor/scalar ratio $r$ and running $\alpha$. Solid contours show constraints from Planck + WMAP Polarization + Lensing + ACT + SPT + BICEP2. Dashed contours show constraints from Planck + WMAP Polarization + Lensing + BAO + BICEP2. The pivot scale is $k_\star = 0.05 h\ {\rm MpC}^{-1}$.
}
\label{fig:r_alpha}
\end{figure}

\begin{figure}
\includegraphics[keepaspectratio,width=1.00\columnwidth,height=0.40\textheight]{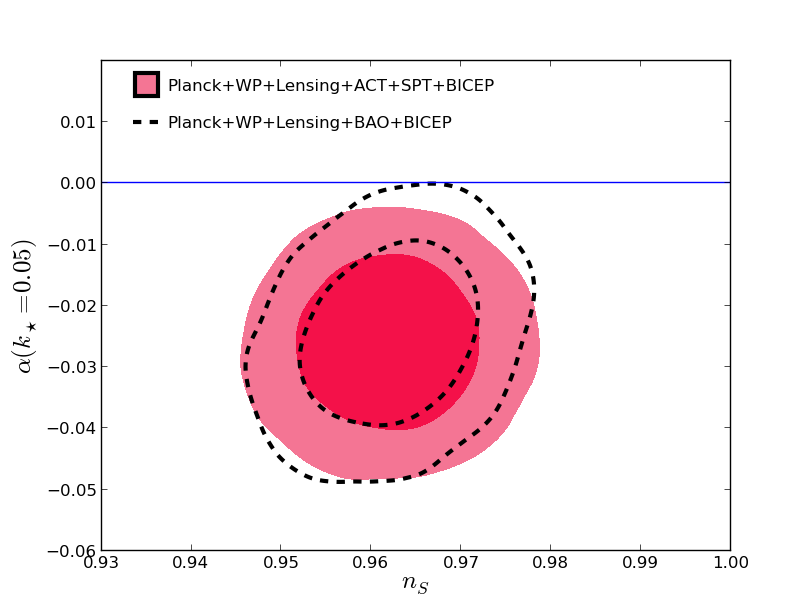}
\caption{Constraints on the scalar spectral index $n_{\rm S}$ and running $\alpha$. Solid contours show constraints from Planck + WMAP Polarization + Lensing + ACT + SPT + BICEP2. Dashed contours show constraints from Planck + WMAP Polarization + Lensing + BAO + BICEP2. The pivot scale is $k_\star = 0.05 h\ {\rm MpC}^{-1}$.}
\label{fig:n_alpha}
\end{figure}

While a simple monomial potential leading to eternal inflation is consistent with the Planck and BICEP2 data, there is tension at approximately the $2\sigma$ level between the Planck and BICEP2 constraints on the amplitude of tensor perturbations \cite{Smith:2014kka}. The tension arises from the anomalously low signal in the temperature anisotropy as observed by Planck on large angular scales \cite{Ade:2013nlj}. The BICEP2 constraint on the tensor amplitude exacerbates the anomaly, since tensor modes {\it also} contribute to the temperature anisotropy on exactly the scales where the Planck CMB power is anomalously low: If, as suggested by BICEP2, as much as 20\% of the large-scale temperature anisotropy is from tensors, this requires even more drastic suppression of curvature perturbations than in the zero-tensor case. This tension can be alleviated by assuming that the shape of the curvature power spectrum is itself scale-dependent, {\it i.e.} by including running of the power spectrum, 
\begin{equation}
\alpha \equiv \frac{d n_{\rm S}}{d \ln{k}},
\end{equation}
where $n_{\rm S}$ is the spectral index of curvature perturbations,
\begin{equation}
n_{\rm S} \equiv \frac{d \ln{P(k)}}{d \ln{k}}.
\end{equation}
Other possibilities for resolving the tension include an extra neutrino species \cite{Giusarma:2014zza,Zhang:2014dxk,Zhang:2014nta,Dvorkin:2014lea} features in the tensor power spectrum \cite{Contaldi:2014zua,Miranda:2014wga}, early Dark Energy \cite{Xu:2014laa}, a non-Bunch-Davies initial state \cite{Boyanovsky:2006qi,Ashoorioon:2014nta}, isocurvature perturbations \cite{Kawasaki:2014fwa}, or a rapid phase transition during inflation \cite{Miranda:2014wga}. Here we focus on the possibility of running. If running is included in a fit to the Planck + BICEP2 data, it is favored at approximately the 95\% confidence level \cite{Gong:2014cqa,Okada:2014lxa,McDonald:2014kia,Hazra:2014aea,Hu:2014aua,Cheng:2014bta}. Figures \ref{fig:r_alpha} and \ref{fig:n_alpha} show joint constraints on the tensor/scalar ratio $r$, the scalar spectral index $n_{\rm S}$, and the running $\alpha$ for Planck and BICEP2. The constraints are generated using the COSMOMC Markov Chain Monte Carlo code \cite{Lewis:2002ah}, marginalizing over a eight-parameter data set with flat priors: 
\begin{itemize}
\item{Dark Matter density $\Omega_{\rm M} h^2$.}
\item{Baryon density $\Omega_{\rm b} h^2$.}
\item{Reionization optical depth $\tau$.}
\item{The angular size  $\theta$ of the sound horizon at decoupling.}
\item{Scalar spectrum normalization $A_{\rm S}$.}
\item{Tensor/scalar ratio $r$.}
\item{Scalar spectral index $n_{\rm S}$.}
\item{Running $\alpha$.}
\end{itemize}
The fit assumes a flat universe $\Omega_{\rm b} + \Omega_{\rm M} + \Omega_{\rm \Lambda} = 1$, with Cosmological Constant Dark Energy, $\rho_{\Lambda} = {\rm const.}$ Convergence is determined via a Gelman and Rubin statistic. Auxiliary data sets used are WMAP polarization (WP), in combination with the Atacama Cosmology Telescope (ACT) / South Pole Telescope (SPT) CMB measurements (solid contours in figures), and Baryon Acoustic Oscillation (BAO) data from Sloan Digital Sky Survey Data Release 9 \cite{Ahn:2012fh}, the 6dF Galaxy Survey \cite{Jones:2009yz}, and the WiggleZ Dark Energy Survey \cite{Blake:2011en} (dashed contours). The pivot scale is taken to be $k_\star = 0.05 h\ {\rm MpC}^{-1}$. 

We see that a running power spectrum is favored at roughly 95\% confidence, a result which can be considered significant from the standpoint of Bayesian evidence \cite{Abazajian:2014tqa}. In this paper, we consider the significance of this result for the hypothesis of eternal inflation. 

\section{Eternal Inflation and the Curvature Perturbation}
\label{sec:eternal}

So-called ``eternal'' inflation occurs when quantum fluctuations in the inflaton field dominate over the classical field evolution, where the amplitude of quantum fluctuations in the inflaton field during inflation is
\begin{equation}
\delta\phi_Q \equiv \left\langle \delta \phi^2\right\rangle^{1/2} = \frac{H}{2 \pi}.
\end{equation}
For eternal inflation to occur, this quantum fluctuation amplitude must be larger than the classical field variation over approximately a Hubble time,
\begin{equation}
\delta\phi_C = \frac{\dot\phi}{H}.
\end{equation}
Therefore, the condition for eternal inflation can be written
\begin{equation}
\frac{\delta\phi_Q}{\delta\phi_C} = \frac{H^2}{2 \pi \dot\phi} > 1.\label{eq:eternal}
\end{equation}
We note that the fraction (\ref{eq:eternal}) is identical to the amplitude of the curvature perturbation for modes crossing the horizon during inflation,
\begin{equation}
P\left(k\right) = \frac{H^2}{2 \pi \dot\phi}.
\end{equation}
The curvature perturbation is simply the amplitude of quantum fluctuations in the inflaton in units of the field variation in a Hubble time! Accordingly, the condition for eternal inflation is that the curvature perturbation amplitude must exceed unity \cite{Goncharov:1987ir,Guth:2007ng},
\begin{equation}
P\left(k\right) > 1.
\end{equation}
In the next section, we show that for sufficiently negative running of the scalar spectral index, the scalar perturbation amplitude always remains below unity and eternal inflation never occurs. 

\section{An upper bound on running}
\label{sec:limit}

The power spectrum for curvature perturbations can be written in terms of the scalar spectral index $n_{\rm S}$ and the running $\alpha$ as
\begin{equation}
P(k) = P_\star \left(\frac{k}{k_\star}\right)^{n_{\rm S} - 1 + \alpha \ln\left(k/k_\star\right) + \cdots},
\end{equation}
where $k_\star$ is a pivot scale, which we take to be a scale comparable to CMB observations, $k_\star = 0.05\ h\ {\rm MpC^{-1}}$, so that a Planck + BICEP + BAO limit gives $P_\star = 2 \times 10^{-9}$. First consider the case of no running, $\alpha = 0$, so that 
\begin{equation}
P(k) = P_\star \left(\frac{k}{k_\star}\right)^{n_{\rm S} - 1}.
\end{equation}
The Planck satellite measurement of the CMB indicates that the spectral index is {\it red}, {\it i.e.} $n_{\rm S} - 1 < 0$, with a 95\% confidence limit of approximately $0.94 < n_{\rm S} < 0.98$. Taking the shallowest power spectrum allowed by Planck, we can put a lower-bound on the amplitude of the power spectrum on scales outside our current horizon size, $k \ll k_\star$, of
\begin{equation}
P\left(k\right) > P_\star \left(\frac{k_\star}{k}\right)^{0.02},\ k < k_\star.
\end{equation}
For a constant red spectral index, we then see that eternal inflation is inevitable in the limit $k \rightarrow 0$, with $P\left(k\right)$ exceeding unity when
\begin{equation}
\frac{k}{k_\star} < P_\star^{1/0.02} \simeq 10^{-435},
\end{equation}
or roughly $N \simeq - \ln{10^{-435}} \simeq 1000$ e-folds before scales of order the current horizon size exited the horizon during inflation. Therefore, for constant spectral index $n_{\rm S} - 1 < -0.02$, eternal inflation is guaranteed as long as inflation continues for at least $1000$ e-folds.

We now consider the case of constant running,
\begin{equation}
P(k) = P_\star \left(\frac{k}{k_\star}\right)^{n_{\rm S} - 1 + \alpha \ln\left(k/k_\star\right)}.
\end{equation}
The Planck + BICEP limit on running of the spectral index is approximately $0 \geq \alpha > -0.05$, so that negative running is favored, and positive running is inconsistent with the data at 95\% confidence. Note that negative running means the spectral index gets {\it redder} on small scales $k \rightarrow \infty$, and {\it bluer} on large scales, $k \rightarrow 0$. For constant negative running $\alpha < 0$, the spectral index for $k \ll k_\star$ will eventually exceed unity, $n - 1 > 0$. We now show that this is sufficient to prevent eternal inflation as long as $\alpha$ is sufficiently negative. If eternal inflation is to be evaded, this implies an upper bound on the curvature perturbation spectrum $P\left(k\right) < 1$ on large scales, or
\begin{equation}
\ln{P\left(k\right)} = \ln{P_\star} + \left[n_{\rm S} - 1 + \alpha \ln{\left(\frac{k}{k_\star}\right)}\right] \ln{\left(\frac{k}{k_\star}\right)} < 0,
\end{equation}
for all $k < k_\star$. The curvature power spectrum will have an extremum at
\begin{equation}
\frac{d \ln{P\left(k\right)}}{d \ln{k}} = n_{\rm S} - 1 + 2 \alpha \ln{\left(\frac{k}{k_\star}\right)} = 0.
\end{equation}
Solving for the wavenumber $k$ gives
\begin{equation}
\ln{\left(\frac{k_{\rm max}}{k_\star}\right)} = \frac{1 - n_{\rm S}}{2 \alpha}.
\end{equation}
This extremum is guaranteed to be a maximum as long as the running $\alpha$ is negative, since
\begin{equation}
\frac{d^2 \ln{P\left(k\right)}}{d \left(\ln{k}\right)^2} = \alpha < 0.
\end{equation}
For eternal inflation to be prevented, it is then sufficient that the maximum of the curvature power spectrum be less than unity, or
\begin{equation}
\ln{P\left(k_{\rm max}\right)} = \ln{P_\star} - \frac{\left(1 - n_{\rm S}\right)^2}{4 \alpha} < 0.
\end{equation}
This is equivalent to an {\it upper bound} on the running $\alpha$ of 
\begin{equation}
\alpha < \frac{\left(1 - n_{\rm S}\right)^2}{4 \ln{P_\star}}.
\end{equation}
From the CMB limits $P_{\star} \simeq 2 \times 10^{-9}$ and $1 - n_{\rm S} < 0.06$, we then have an upper bound on the running
\begin{equation}
\alpha < - 4 \times 10^{-5}.\label{eq:alphalb}
\end{equation}
For running below this upper bound, the curvature perturbations amplitude remains smaller than unity for all wavenumbers $k$. Thus, even a very weak negative running, consistent with slow-roll inflation, is sufficient to prevent eternal inflation from ever occurring. 

We have assumed that the running of the spectral index is constant, {\it i.e.} that there is no running-of-running, running-of-running-of-running, and so on. There is, however, no guarantee that the contributions from higher-order terms in the series
\begin{equation}
\left(n_{\rm S} - 1\right) + \alpha \ln{\left(\frac{k}{k_\star}\right)} + \beta \left[\ln{\left(\frac{k}{k_\star}\right)}\right]^2 + \cdots
\end{equation}
do not become important in the limit $k \rightarrow 0$. It is evident that, as long as the higher-order terms are themselves negative, eternal inflation will still never occur: $\alpha$ need not be constant to suppress eternal inflation, it must simply be negative. In addition, negative running of the spectral index also implies an eventual breakdown in slow roll at very large scales, indicating a finite total duration of inflation. We can estimate the total number of e-folds of inflation by noting that $d N = d \ln{a} \simeq d \ln{k}$, where $a(t)$ is the scale factor. Then, a breakdown of slow roll will occur when
\begin{equation}
\left\vert \int{\frac{d n_{\rm S}}{d\ln{k}} d\ln{k}} \right\vert \sim 1,
\end{equation}
which gives $N \sim \left\vert \Delta \ln{k}\right\vert \sim 1 / \left\vert\alpha\right\vert$. For running $\alpha \sim -0.05$, near the outer limit of the Planck + BICEP allowed region, this means a very rapid breakdown of slow roll, in about 20 e-folds of inflation. However, for more moderate running, a lengthy period of inflation is still possible, of order $10^{4}$ e-folds for $\alpha \sim -10^{-4}$. Inflation can continue for an extended period without the onset of eternal inflation. However, inflation in such cases is still of finite duration. To avoid eternal inflation, it is sufficient that higher-order terms be subdominant during the {\it finite} period of inflation, for example
\begin{equation}
\alpha > \beta \left\vert\Delta\ln{k}\right\vert,
\end{equation}
or 
\begin{equation}
\beta < \alpha^2.
\end{equation}
This is a somewhat stronger condition than the assumption of slow roll, since in the slow roll approximation, the spectral index is first-order in slow roll parameters, $n - 1 \sim {\mathcal O}\left(\epsilon,\eta\right)$, $\alpha \sim {\mathcal O}\left(\epsilon^2,\epsilon\eta,\ldots\right)$, $\beta \sim {\mathcal O}\left(\epsilon^3,\ldots\right)$. This condition is sufficient, but not necessary: eternal inflation may still be suppressed even if the running becomes positive, as long as the curvature perturbation remains below unity.

What does the lower bound (\ref{eq:alphalb}) imply about the form of single-field inflationary potentials? In terms of slow roll parameters, we can write
\begin{equation}
n_{\rm S} - 1 = 2 \eta - 6 \epsilon,
\end{equation}
and
\begin{equation}
\alpha = -2 \xi + 16 \epsilon \eta - 24 \epsilon^2,
\end{equation}
where
\begin{eqnarray}
\epsilon &=& \frac{M_{\rm P}^2}{2} \left(\frac{V'}{V}\right)^2\cr
\eta &=& M_{\rm P}^2 \left(\frac{V''}{V}\right)\cr
\xi &=& M_{\rm P}^4 \left(\frac{V' V'''}{V^2}\right).
\end{eqnarray}
For eternal inflation to be suppressed, the spectral index must change from red ($n < 1$) on small scales to blue ($n > 1$) on large scales. Since $\epsilon$ is positive-definite, a blue spectrum means that $\eta$ must be positive and large relative to $\epsilon$,
\begin{equation}
\eta > 3 \epsilon.
\end{equation}
Taking $V', V'' > 0$, so that the potential becomes large for large field values, we must then have
\begin{equation}
\frac{d \eta}{d\phi} = M_{\rm P}^2 \frac{V'''}{V} - M_{\rm P}^2 \frac{V' V''}{V^2} > 0.
\end{equation}
Therefore, negative running requires a large positive third derivative $V'''$ of the potential, or equivalently a slow-roll parameter $\xi > 2 \epsilon \eta$. A simple example of such a potential is inflation near an inflection point \cite{Choudhury:2014kma},
\begin{equation}
\label{eq:potential}
V\left(\phi\right) = V_0 + \Lambda^3 \phi - m^2 \phi^2 + \mu \phi^3 + \cdots,
\end{equation}
where the constants $\Lambda$, $m$, and $\mu$ all have dimensions of mass. Near the inflection point $\phi = 0$,
\begin{equation}
\eta = M_P^2 \frac{3 \mu\phi -2 m^2 }{V_0},
\end{equation}
which is positive for $\phi > 2 m^2 / 3 \mu$, and negative for $\phi < 2 m^2 / 3 \mu$, so the spectral index evolves from blue ($n_{\rm S} > 1$) to red ($n_{\rm S} < 1$) as the field rolls down the potential. A sufficiently large tensor/scalar ratio can be generated by tuning of the coefficient of the linear term, since near $\phi = 0$,
\begin{equation}
r = 16 \epsilon \simeq 8  \left(\frac{M_{\rm P}\Lambda^3}{V_0}\right)^2. 
\end{equation}
Such inflection point models have been suggested as typical in the string landscape \cite{Enqvist:2010vd,Allahverdi:2006we,Agarwal:2011wm,Cicoli:2013oba,Choudhury:2013jya,Bousso:2013uia,Bousso:2014jca}, albeit typically with very small tensor/scalar ratios arising from the necessity of small field variation on the compactified manifolds typical in string theory. A thorough dynamical analysis of inflection-point inflation can be found in Ref. \cite{Downes:2012gu}.  Similar models have been proposed from a strictly phenomenological viewpoint, with suppression of the curvature perturbation on large scales arising from a period of fast-roll scalar field evolution \cite{Contaldi:2003zv,Lello:2013awa,Handley:2014bqa,Hazra:2014jka}. Other possibilities include an early superinflationary phase \cite{Liu:2013iha}, an early non-inflationary phase \cite{Hirai:2005pg,Powell:2006yg,Nicholson:2007by,Ramirez:2011kk,Ramirez:2012gt}, double inflation \cite{Kawasaki:2003zv}, a curvaton \cite{Sloth:2014sga}, punctuated inflation \cite{Jain:2009pm,Jain:2008dw}, or other new physics \cite{BasteroGil:2003bv,Buchel:2004df}. The possibilites for model-building extend well beyond the simple potential (\ref{eq:potential}). 

\section{Conclusions}
\label{sec:conclusions}

In this paper, we consider the viability of eternal inflation in light of the results from the Planck and BICEP2 observations of the Cosmic Microwave Background. Current data weakly favor nonzero running of the scalar spectral index $n_{\rm S}$, mostly as a result of the suppressed scalar power observed on large angular scales in the Planck data. The suppression of low-$\ell$ modes in the CMB, compared to expectations from the standard $\Lambda$-CDM cosmology, may be due either to negative running, or may have another more exotic origin. In this paper, we consider eternal inflation in a scenario with nonzero running, and show that a negative running of the scalar spectral index on superhorizon scales serves to suppress eternal inflation. Assuming a constant running, we derive an upper bound 
\begin{equation}
\alpha < - 4 \times 10^{-5}.
\end{equation}
For running below this bound, the primordial power spectrum is less than unity on all scales larger than the current horizon, and eternal inflation is prevented. In a more realistic case where higher-order terms such as running-of-running become significant, it is still the case that, as long as the curvature perturbation remains smaller than order unity, eternal inflation does not occur. In single-field inflationary models, negative running eventually results in a breakdown of slow roll and therefore a finite duration for inflation. We show that negative running is consistent with as many as $10^4$ e-folds of inflation, without the onset of eternal inflation, in contrast to the case of no running, for which eternal inflation will occur given around $1000$ e-folds of inflation. 

Can we really know what occurs very early in the inflationary epoch? Eternal inflation requires a curvature perturbation spectrum of at least order unity to occur, $P\left(k\right) \geq 1$ (which itself raises concerns about the effect of gravitational backreaction \cite{Martinec:2014uva}). It is therefore clear that the portion of the potential that produces observable density fluctuations ({\it i.e.} around 60 e-folds before the end of inflation) cannot give rise to eternal inflation, since CMB normalization requires $P\left(k\right) \sim 10^{-10}$. Eternal inflation occurs on the portion of the potential where the inflaton field rolls prior to producing these perturbations, which corresponds to length scales larger than our current horizon size. For example, in an $m^2 \phi^2$ potential, with $m\sim 10^{13}$ GeV, eternal inflation only takes place high up in the potential, at $\phi > 100 M_{\rm P}$. Since we do not have observational access to superhorizon length scales, and therefore the physics of very early stages of inflation, any conclusion we might reach contains an inherent element of speculation: It may be that such high regions of the potential are never probed, for example in the case of non-negligible spatial curvature \cite{Kleban:2012ph}. In this paper, we have not {\it proven} that eternal inflation does not occur. We have argued that it is not inevitable, even in single-field inflation, and current data in fact hint that we may be in a situation where eternal inflation is suppressed, even on far super-horizon scales.

\section*{Acknowledgments}

KF acknowledges the support of the DOE under grant DOE-FG02-95ER40899 and the Michigan Center for Theoretical Physics at the University of Michigan. WHK is supported by the National Science Foundation under grant NSF-PHY-1066278. We thank Alejandro Lopez for helpful conversations. This work was performed in part at the University at Buffalo Center for Computational Research. 

\bibliography{paper}

\end{document}